\documentclass[a4paper]{jpconf}
\usepackage{graphicx}
\usepackage{citesort}

\begin{document}

\title{Results from NEMO-3.}
\author{Robert L. Flack on behalf of the NEMO-3 collaboration.}

\address{Department of Physics and Astronomy, University College London, Gower Street, LONDON, WC1E 6BT, UK}

\ead{robflack@hep.ucl.ac.uk}

\begin{abstract}
The NEMO-3 experiment is located in the Modane Underground Laboratory
(LSM) and has been taking data since 2003 with seven isotopes. It is
searching for the double beta decay process with two
($2\nu\beta\beta$) or zero ($0\nu\beta\beta$) neutrinos emitted in the
final state.  Precision measurements of the half-life of the isotopes
due to $2\nu\beta\beta$ decay have been performed and new results for
$\rm T_{1/2}^{2\nu}(^{96}Zr) = [2.3 \pm 0.2 (stat) \pm 0.3
(syst)]\cdot10^{19}y$, $\rm T_{1/2}^{2\nu}(^{48}Ca)=[4.4
^{+0.5}_{-0.4} (stat) \pm 0.4(syst)]\cdot10^{19}y$ and $\rm
T_{1/2}^{2\nu}(^{150}Nd)=[0.920^{+0.025}_{-0.022} (stat) \pm 0.072
(syst)] \cdot10^{19}y$ are presented here.  Measurements of this
process are important for reducing the uncertainties on the nuclear
matrix elements. No evidence for $0\nu\beta\beta$ decay has been found
and a 90\% Confidence Level lower limit on the half-life of this
process is derived. From this an upper limit can be set on the
effective Majorana neutrino mass using the most recent nuclear matrix
elements (NME) calculations.
\end{abstract}

\section{Introduction}
The objective of the NEMO~3 experiment is to search for the double
beta decay process with two ($2\nu\beta\beta$ decay) or zero
($0\nu\beta\beta$ decay) neutrinos in the final state in the seven
different $\beta\beta$ isotopes listed in Table~\ref{tab:t12}.  The
experimental search for $0\nu\beta\beta$ decay is of major importance
in particle physics. If this process is observed then it will reveal
the Majorana nature of the neutrino ($\nu \equiv\bar{\nu}$) and may
allow an access to the absolute neutrino mass scale.

The process $0\nu\beta\beta$ decay also violates the principle of lepton number
conservation and is, therefore, a direct probe for physics beyond the
standard model.  In the case of the neutrino-mass mechanism the
$0\nu\beta\beta$ decay rate can be written as
\begin{equation}
 [T_{1/2}^{0\nu}(A,Z)]^{-1} = \langle m_{\nu}\rangle^2 \cdot |M^{0\nu}(A,Z)|^2 \cdot G^{0\nu}(Q_{\beta\beta},Z) , 
\label{eq:nme} 
\end{equation} 
where $\langle m_{\nu}\rangle$ is the effective neutrino mass,
$M^{0\nu}$ is the nuclear matrix element (NME), and $G^{0\nu}$ is the
kinematical factor proportional to the transition energy to the fifth
power, $Q_{\beta\beta}^5$.

The $2\nu\beta\beta$ decay process is a rare second order weak interaction
process. The accurate measurement of its rate of decay is important since it
constitutes the ultimate background in the search for
$0\nu\beta\beta$ decay signal and is a valuable input for the
theoretical calculations of the NME.

\section{The NEMO~3 detector}\label{subsec:detector}
The detector~\cite{nemo3} is located in the LSM in the Frejus tunnel
at the depth of 4800 m w.e.  It is cylindrical with thin source foils
($\rm \sim 50\,mg/cm^2$) situated in the middle of the tracking volume
surrounded by the calorimeter.  The source foils are composed of
almost 10\,kg of enriched $\beta\beta$ isotopes, listed in
Table~\ref{tab:t12}.  Observation of the $\beta\beta$ decay is
accomplished by fully reconstructing the tracks of the two electrons
and measuring their energy.  A tracking chamber containing 6180 open
drift cells, operating in the Geiger mode, provides a vertex
resolution of about 1\,cm and a 25\,G magnetic field is used to curve
the tracks for charge identification.  A calorimeter consisting of
1940 plastic scintillator blocks with photomultiplier readout gives an
energy resolution of 14 --17\%/$\rm \sqrt{E\,(MeV)}$ FWHM. A time
resolution of 250\,ps allows excellent suppression of the external
background due to electrons crossing the detector.  The detector is
capable of identifying e$^-$, e$^+$, $\gamma$ and $\alpha$ particles
and allows good discrimination between signal and background events.
The detector is covered by two layers of shielding against external
$\gamma$ rays and neutrons.

\section{Event selection and background model}
The $\rm \beta\beta$ events are selected by requiring two
reconstructed electron tracks with a curvature corresponding to the
negative charge, originating from a common vertex in the source foil.
The energy of each electron measured in the calorimeter should be
higher than 200\,keV.  Each track must be incident upon a separate
scintillator block and the time difference of the two PMT signals is
compared to the estimated time as if the two electrons had originated
from a common vertex.

The background can be classified in three groups: external one from
incoming $\rm \gamma$, radon inside the tracking volume and internal
radioactive contamination of the source.  All three were estimated
from the NEMO-3 data with events of various topologies.  In
particular, radon was measured with e$\rm\gamma$ and e$\rm\alpha$
events and internal $\rm^{214}$Bi with e$\rm\alpha$. The
e$\rm\gamma$$\rm\gamma$, and e$\rm\gamma$$\rm\gamma$$\rm\gamma$ events
are used to measure the $\rm^{208}$Tl activity requiring the detection
of the 2.615 MeV $\rm\gamma$-ray typical of the $\rm^{208}$Tl
$\rm\beta$ decay.  Single electron events are used to measure the foil
contamination by $\rm\beta$-emitters.  The external background is
measured with the events with the detected incoming $\rm\gamma$-ray
producing an electron in the source foil. The external background is
checked with two-electron events originating from pure copper and
natural tellurium foils.  Measurements performed with an HPGe detector
and radon detectors are used to verify the results.

\section{Results}
Measurements of the $\rm 2\nu\beta\beta$ decay half-lives have been
performed for the seven isotopes available in NEMO-3 and for
completeness previous results as well as new results are given in
Table~\ref{tab:t12}.  New preliminary results based on higher
statistics than used previously are presented here for the three isotopes,
$\rm ^{48}Ca$, $\rm ^{96}Zr$ and $\rm ^{150}Nd$ .

\begin{table}[hbt]
\caption{NEMO-3 results of $\rm 2\nu\beta\beta$ decay half-life measurements for seven isotopes. \label{tab:t12}}
\begin{center}
\begin{tabular}{ c|c|c|c|c }
\hline
Isotope& Mass (g) & Q$\rm_{\beta\beta}$ (keV) & Signal/Background & T$\rm_{1/2}$ [$\rm10^{19}$ years]\\
\hline
$\rm ^{100}Mo$ &6914& 3034 & 40   & 0.711 $\rm\pm$ 0.002 (stat) $\pm$ 0.054 (syst)~\cite{prl}\\
$\rm ^{82}Se$  &932 & 2995 & 4    & 9.6 $\pm$ 0.3 (stat) $\pm$ 1.0 (syst)~\cite{prl}\\
$\rm ^{116}Cd$ & 405& 2805 & 7.5  & 2.8 $\pm$ 0.1 (stat) $\pm$ 0.3 (syst)~\cite{Soldner-Rembold}\\
$\rm ^{150}Nd$ &37.0& 3367 & 2.8  & $0.920 ^{+0.025}_{-0.022} $(stat) $\pm$ 0.072 (syst)\\
$\rm ^{96}Zr$  &9.4 & 3350 & 1.   & 2.3 $\pm$ 0.2 (stat) $\pm$ 0.3 (syst)\\
$\rm ^{48}Ca$  &7.0 & 4272 & 6.8  & $4.4 ^{+0.5}_{-0.4} $(stat) $\pm$ 0.4 (syst)\\
$\rm ^{130}Te$ &454 & 2529 & 0.25 & 76 $\pm$ 15 (stat) $\pm$ 8 (syst)~\cite{Soldner-Rembold}\\
 \hline
\end{tabular}
\end{center}
\end{table}

\begin{figure}[htb]
\begin{center}
\includegraphics[width=0.41\textwidth,height=4.3cm]{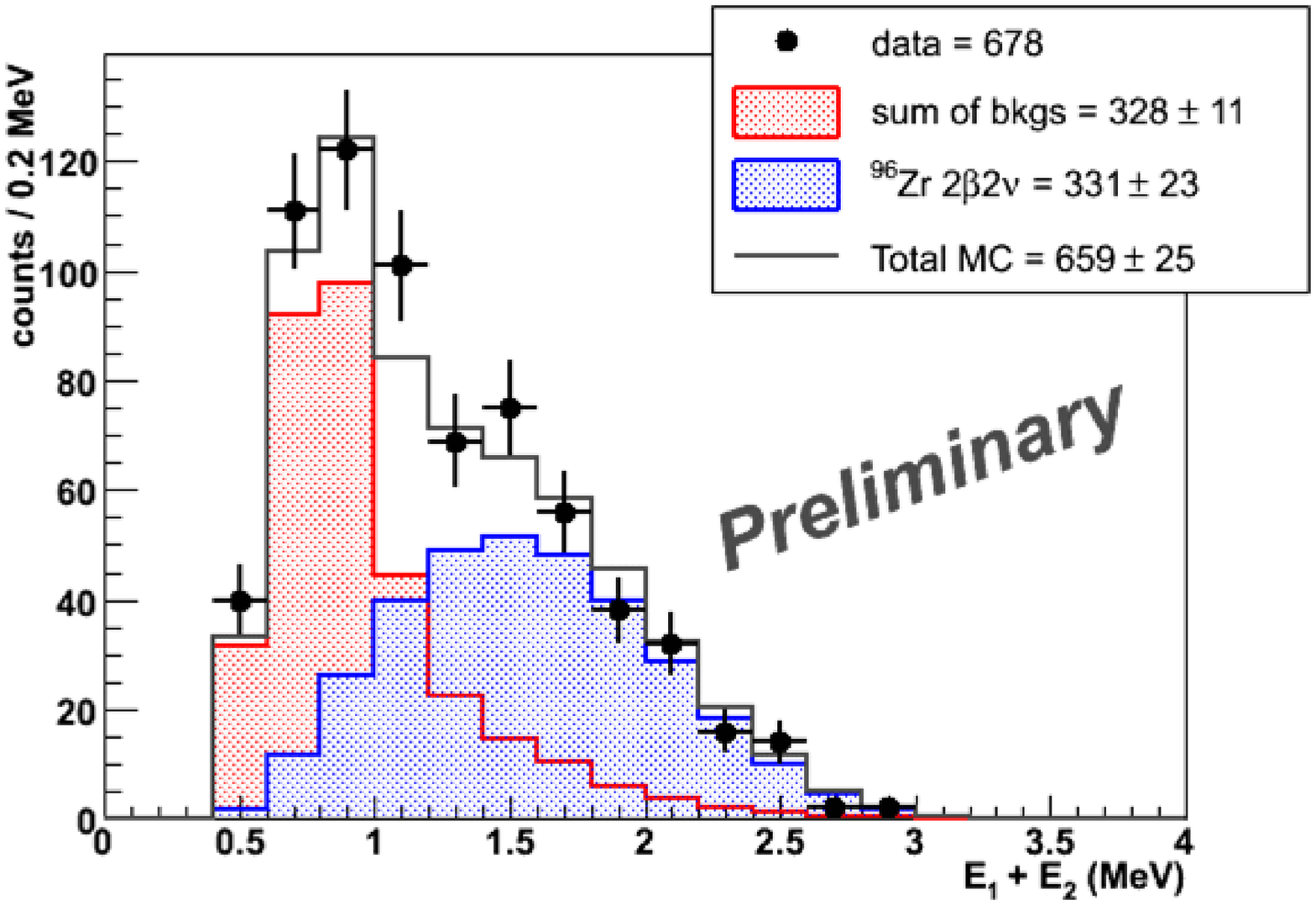}
\includegraphics[width=0.41\textwidth,height=4.3cm]{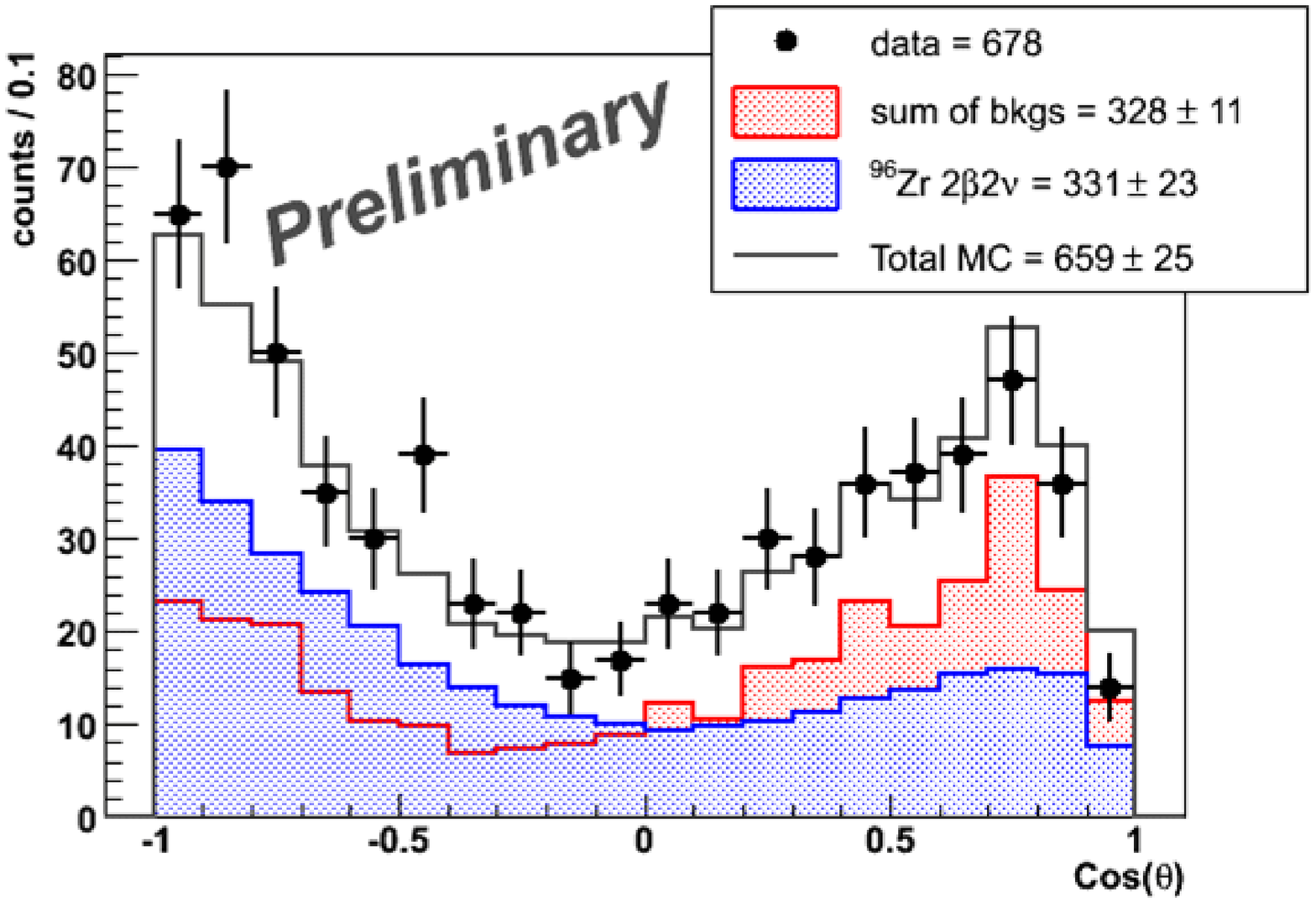}
\end{center}
\caption{The sum of the energies of the two electrons and their angular separation
for two-electron events for $^{96}$Zr.}
\label{fig:bb_zr96}
\end{figure}

The measurement of the $\rm ^{96}Zr$ half-life was performed using the
data collected for 925 days. A total of 678 events was selected,
with an expectation of 328 background events. The largest background
contribution is due to the internal $\rm ^{40}K$ contamination of the
sample. The distributions of the sum of the energies of the two electrons and their
angular separation are shown in Fig.~\ref{fig:bb_zr96},
demonstrating good agreement between the data and the Monte Carlo
simulation.  The $\rm 2\nu\beta\beta$ efficiency is estimated to be
7.6\%. The measured half-life is $\rm T_{1/2}^{2\nu}(^{96}Zr) = [2.3 \pm 0.2 (stat) \pm 0.3 (syst)]\cdot10^{19}y$.

\begin{figure}[htb]
\begin{center}
\includegraphics[width=0.41\textwidth,height=4.3cm]{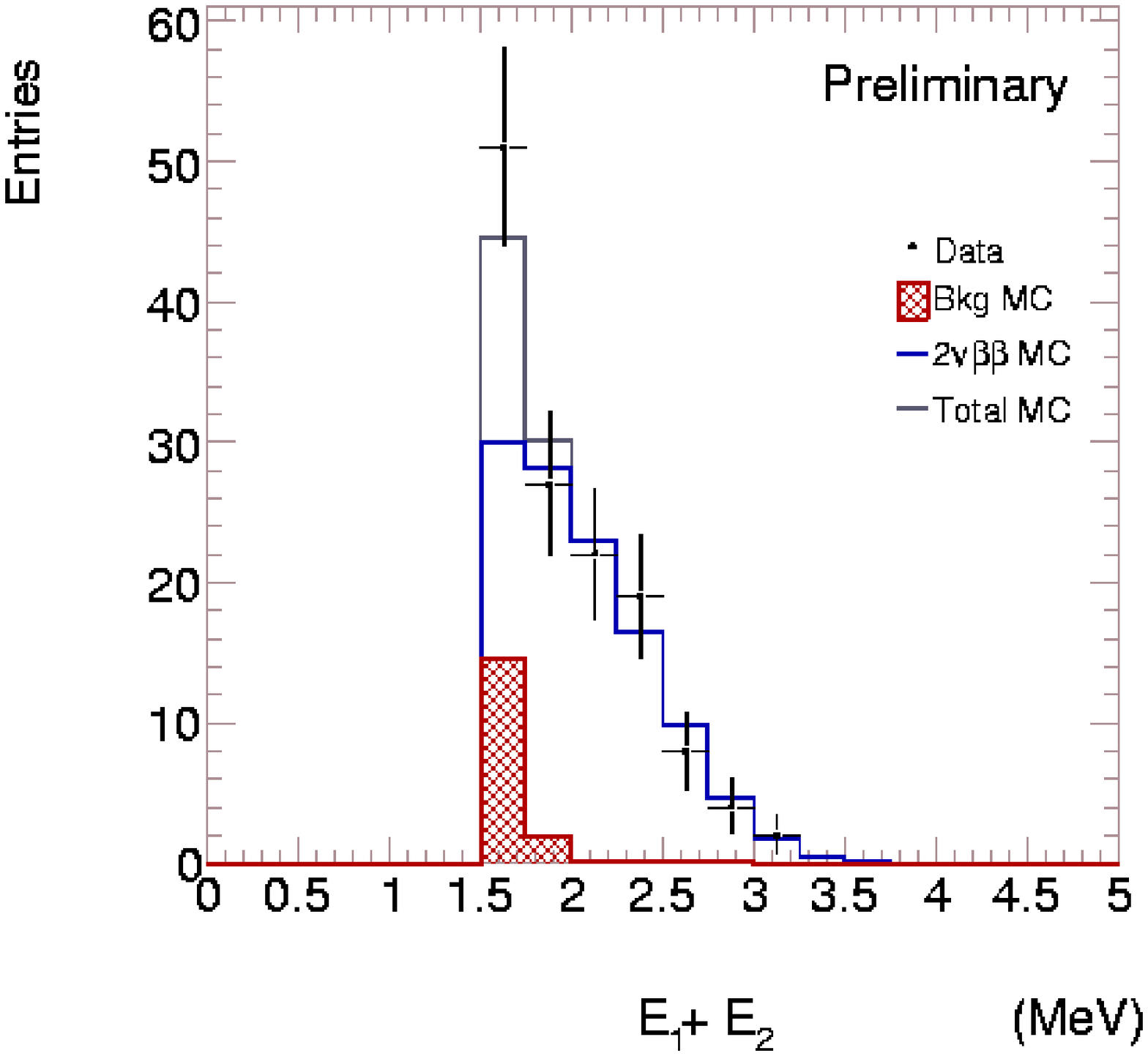}
\includegraphics[width=0.41\textwidth,height=4.3cm]{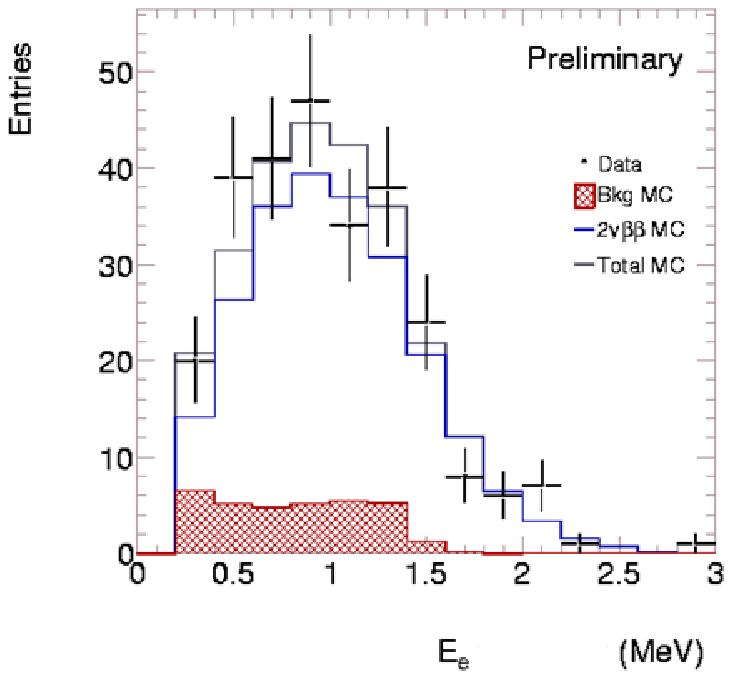}
\end{center}
\caption{The energy sum and single-electron energy spectra  
for two-electron events from $^{48}$Ca.}
\label{fig:bb_ca48}
\end{figure}

The measurement of the $\rm ^{48}Ca$ half-life was performed using the
data collected for 943 days. A total of 133 events was selected, with
17 background events expected. The two-electron energy sum
distribution and single-electron energy spectrum are presented in
Fig.~\ref{fig:bb_ca48}.  The $\rm ^{48}$Ca sample is known to be
contaminated with $\rm ^{90}$Sr (T$\rm _{1/2}$=28.79\,y, $\rm
Q_{\beta}$=0.546\,MeV). Its daughter $\rm ^{90}$Y (T$\rm
_{1/2}$=3.19\,h, $\rm Q_{\beta}$=2.282\,MeV) is the major background
source in this case. An activity of approximately 1700\,mBq/kg was
measured for $\rm ^{90}$Y using single-electron events.  Both $\rm
^{90}$Sr and $\rm ^{90}$Y are essentially pure $\rm \beta^-$ emitters
and imitate $\rm \beta\beta$ events through M\"oller scattering. To
suppress this background contribution, events with the energy sum
greater than 1.5 MeV and $\rm cos(\Theta_{ee})<0$ are selected.  The
$\rm 2\nu\beta\beta$ efficiency is 3.3\%, and the measured half-life
is $\rm T_{1/2}^{2\nu}(^{48}Ca)=[4.4 ^{+0.5}_{-0.4} (stat) \pm
0.4(syst)]\cdot10^{19}y$.

\begin{figure}[htb]
\begin{center}
\includegraphics[width=0.41\textwidth,height=4.3cm]{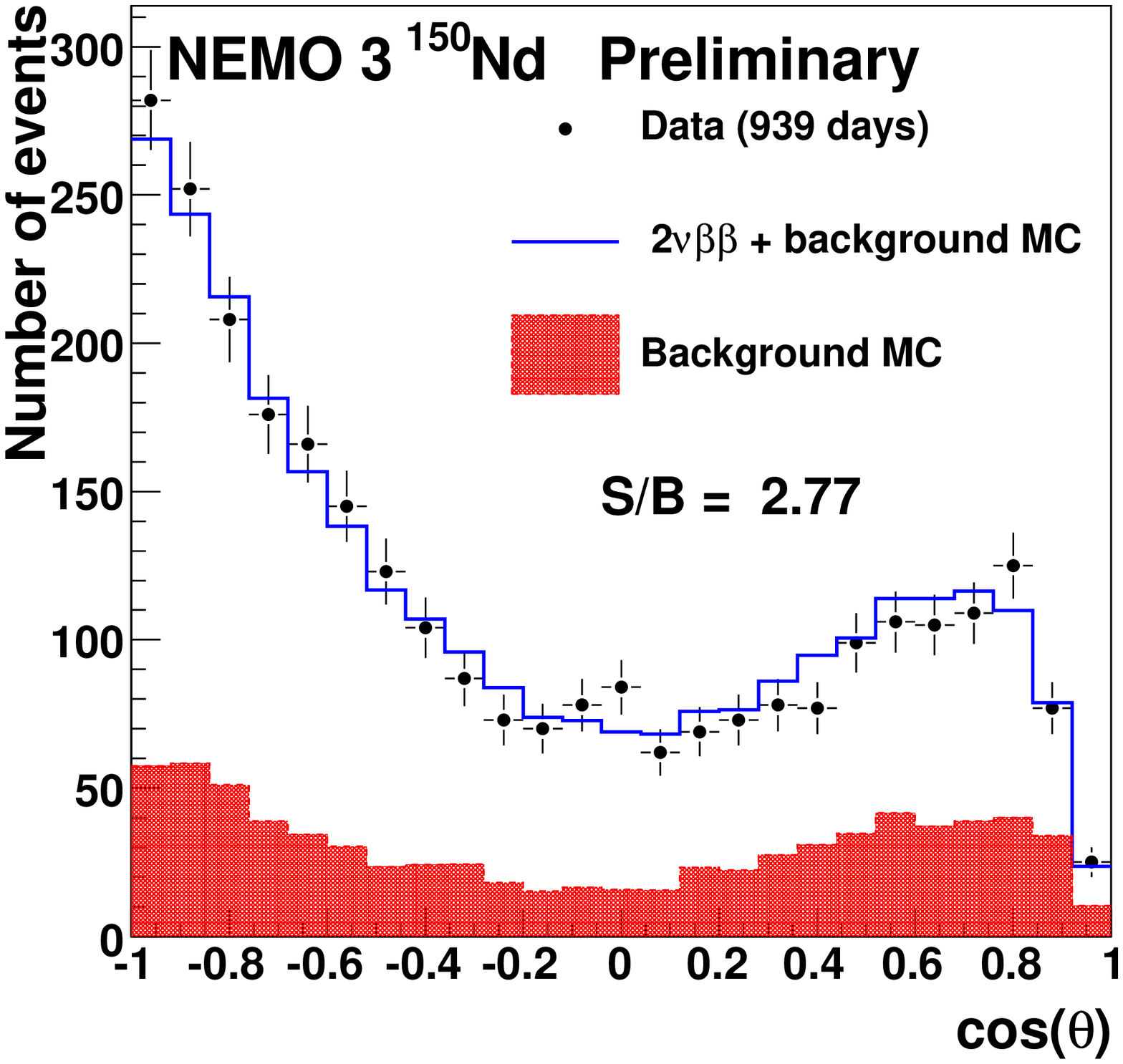}
\includegraphics[width=0.41\textwidth,height=4.3cm]{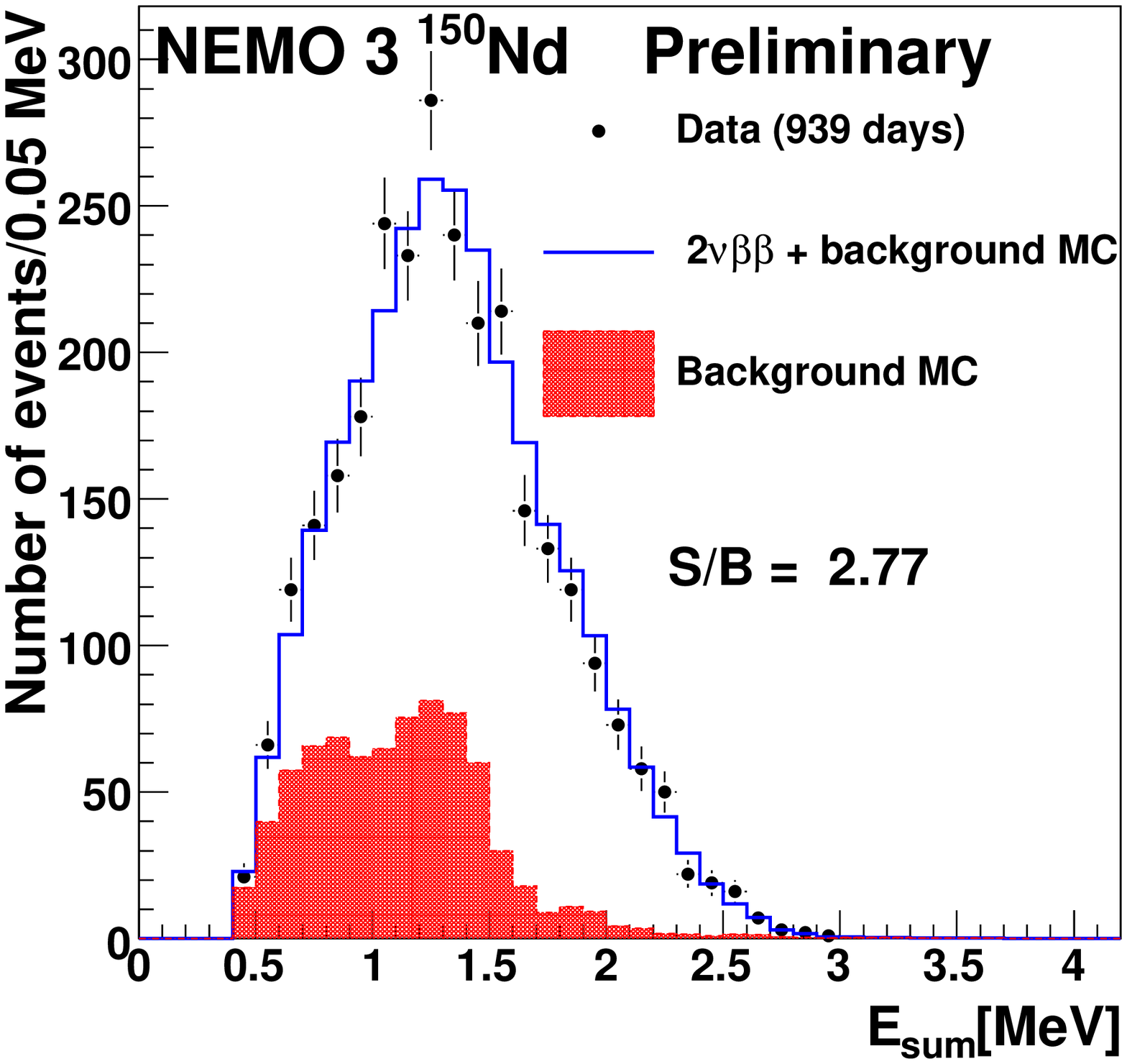}
\end{center}
\caption{The distributions of the angular separation and energy sum spectra for two-electron events from $^{150}$Nd.}
\label{fig:bb_nd150}
\end{figure}

The measurement of the $\rm ^{150}Nd$ half-life was performed using the
data collected within 939 days. A total of 2853 events was selected,
with 765 background events expected.  The distributions of the
sum of the energies of the two electrons and their angular separation are shown
in Fig.~\ref{fig:bb_nd150}, demonstrating good agreement between the data
and the Monte Carlo simulation.  The $\rm 2\nu\beta\beta$ efficiency is
estimated to be 7.6\% and the half-life is measured to be
$\rm T_{1/2}^{2\nu}(^{150}Nd)=[0.920^{+0.025}_{-0.022} (stat) \pm 0.062 (syst)] \cdot10^{19}y$.

In the case of the mass mechanism, the $0\nu\beta\beta$ decay signal
is expected to be a peak in the energy sum distribution at the
position of the transition energy $\rm Q_{\beta\beta}$ (with a long
non-Gaussian tail due to energy losses in the source foil).  Since no
excess is observed at the tail of the distribution for $^{96}$Zr, see
Fig.~\ref{fig:bb_zr96} (left), nor for $^{48}$Ca,
Fig.~\ref{fig:bb_ca48} (left), limits are set on the neutrinoless
double beta decay T$_{1/2}^{0\nu}$ using the CLs method~\cite{cls}.  A
lower half-life limit is translated into an upper limit on the
effective Majorana neutrino mass $\langle m_{\nu}\rangle$ (see
Eq.~\ref{eq:nme}).  The following results are obtained using the NME
values from~\cite{Kort,Simkovic} for $^{96}$Zr and from~\cite{Caurier} for
$^{48}$Ca:
T$\rm _{1/2}^{0\nu}(^{96}Zr) > 8.6\cdot 10^{21} y$ (90\% C.L.),  $\langle m_{\nu}\rangle < $ 7.4 -- 20.1\,eV 
and T$\rm _{1/2}^{0\nu}(^{48}Ca) > 1.3\cdot 10^{22} y$ (90\% C.L.),  $\langle m_{\nu}\rangle < $ 29.7\,eV.

Similarly a limit has been set for $\langle m_{\nu}\rangle$ for $\rm ^{150}Nd$. It is complicated by whether the deformation of the nucleus is taken into account.  The limit has been calculated for both scenarios:
T$\rm _{1/2}^{0\nu}(^{150}Nd) > 1.8\cdot 10^{22} y$ (90\% C.L.)  with
$\langle m_{\nu}\rangle < $ 1.7 -- 2.4\,eV using QRPA model (no deformation)~\cite{Rodin} and
$\langle m_{\nu}\rangle < $ 4.8 -- 7.6\,eV using SO(3) model (with deformation)~\cite{Hirsch}.

All of these limits should be compared with the one set previously by NEMO-3 of $\langle m_{\nu}\rangle < $ 0.8 -- 1.3\,eV for $\rm ^{100}Mo$~\cite{prl}.

\section{Conclusion}
 No evidence was found for $0\nu\beta\beta$ decay and limits on the
 neutrino mass, $\langle m_{\nu}\rangle$, have been set.  New
 measurements for the half-lives due to $\rm 2\nu\beta\beta$ decay for
 the three isotopes, $\rm ^{48}Ca$, $\rm ^{96}Zr$ and $\rm ^{150}Nd$
 has been presented.  The NEMO-3 experiment continues taking data but will
 eventually be superceded with a new experiment called SuperNEMO. The
 baseline design of SuperNEMO envisages using 100\,kg of $\rm ^{82}Se$ or
 $\rm ^{150}Nd$ to reach a sensitivity to the Majorana neutrino mass
 of 0.05 -- 0.1\,eV.



\end{document}